\documentstyle[aps,prb,preprint,tighten]{revtex}

\begin{document}

\draft
\title{
\large\bf The effect of the annealing temperature on the local distortion of
La$_{0.67}$Ca$_{0.33}$MnO$_3$ thin films}
\author{\rm D. Cao$^1$, F. Bridges$^1$, D.~C. Worledge$^2$, C.~H. Booth$^3$, 
T. Geballe$^2$} 

\address{\rm
$^{(1)}$Department of Physics, University of California Santa
Cruz, Santa Cruz, CA 95064}

\address{\rm
$^{(2)}$Department of Applied Physics, Stanford University, Stanford, CA
94305-4090}

\address{\rm
$^{(3)}$Los Alamos National Laboratory, Los Alamos, NM 87545}

\date{draft: \today}

\maketitle

\begin{abstract}

Mn $K$-edge fluorescence data are presented for thin film
samples (3000~\AA) of Colossal Magnetoresistive (CMR) 
La$_{0.67}$Ca$_{0.33}$MnO$_3$: as-deposited, and post-annealed
at 1000~K and 1200~K. The local distortion is analyzed in terms of three 
contributions: static, phonon, and an extra, temperature-dependent, polaron term.
The polaron distortion is very small for the as-deposited sample and increases
with the annealing temperature. In contrast, the static distortion in the 
samples 
decreases with the annealing temperature. Although the local structure of the 
as-deposited sample shows very little temperature dependence, the 
change in resistivity with temperature is the largest of these three
thin film samples. The as-deposited sample also has the highest 
magnetoresistance (MR), which indicates some other mechanism may also 
contribute to the transport properties of CMR samples. We also discuss the 
relationship between local distortion and the magnetization of the sample.

\end{abstract}

\pacs{Keywords: CMR, XAFS, thin film, annealing}

\narrowtext

\section{Introduction}

The Double Exchange (DE) mechanism\cite{Zener51,Anderson55,deGennes60}
was originally considered to be the main
interaction contributing to the colossal magnetoresistance
(CMR). In the DE model, if the spins of 
two neighboring Mn ions are aligned, then an electron will require less 
energy to hop from one Mn site to another. Consequently, at low temperature, 
the lattice will have ferromagnetic (FM) order such that the total system 
(both local and itinerant sub-systems) has the lowest energy.
Although the DE model can explain many properties of CMR materials,
the magnitude of
the MR calculated from the DE model is much smaller than the actual measured 
MR\cite{Millis95}.
Millis {\it et al.} suggested that local $\it{Jahn-Teller}$
distortions also play an important role in CMR materials, and are needed
to explain the large magnitude of the MR\cite{Millis95} in these materials.

Both X-ray Absorption Fine Structure 
(XAFS)\cite{Booth98a,Booth98b}
and pair-distribution function (PDF) analysis of neutron diffraction 
data\cite{Billinge96,Louca97}, 
have been done to study the local structure of the CMR materials and 
an important relationship between the local distortions
and magnetism in these materials has been found\cite{Booth98a,Booth98b}.
These new experiments investigate the local structure of thin films of
La$_{0.67}$Ca$_{0.33}$MnO$_3$ (LCMO) to 
understand more about this relationship.

Recent experiments show that the 
non-fully annealed thin film samples are oxygen
deficient\cite{Worledge96,Worledge98}; the Curie temperature, T$_c$, the 
saturated magnetization, M$_0$, and the resistivity 
peak temperature, T$_{MI}$, of the 
samples increase with increasing the oxygen stoichiometry, 
while the resistivity of the 
samples decreases\cite{Worledge96,Worledge98,Ju95,Malde98}. 
For the fully
annealed thin film sample, T$_c$ and T$_{MI}$ are almost the same
as those of the bulk material. It is not yet clear what
mechanisms are responsible for suppressing T$_c$, M$_0$ and T$_{MI}$
in the other samples. Our experiments
on transport and magnetization measurements show similar annealing effects
with significant changes in the resistivity and the magnetization.
They also show that a huge MR occurs for the films,
especially for the as-deposited sample at low temperature.

XAFS experiments on these thin film
samples allow us to observe the local structure around the Mn sites, primarily
the local
distortion of the Mn-O bonds. This paper focuses on the $\it{changes}$ in the 
local
structure of CMR thin films that are induced by annealing at different
temperatures. These results may also help us to better understand how annealing
modifies other sample properties.

Diffraction studies of LaMnO$_3$ and CaMnO$_3$ have been carried out
by several groups\cite{Norby95,Mitchell96,Poeppelmeier82}. 
They show that there are three groups of 
Mn-O bonds in LaMnO$_3$ with different lengths: 1.91, 1.97 and 
2.17 \AA{}\cite{Mitchell96}, while in CaMnO$_3$ all bond lengths nearly the 
same at 
1.90 \AA{} \cite{Poeppelmeier82}(variation is within 0.01 \AA{}).
Our previous work has compared our XAFS data with the diffraction results,
and found that the local structure around the Mn site in LCMO CMR samples is 
very similar to the average structure determined by the diffraction data.
More comparison details are shown in reference \onlinecite{Booth98b}.

In Sec. II, we provide a brief description of the samples and some 
experimental details. We present the magnetization and transport property data
in Sec. III and our XAFS results in Sec. IV.
The conclusions are given in Sec. V. 

\section{Samples and Experiments}

The La$_{0.67}$Ca$_{0.33}$MnO$_3$ thin-film samples were deposited on
SrTiO$_3$ substrates using PLD; each film is 3000~\AA{} thick.
See Ref. \onlinecite{Worledge98} 
for additional details. The samples we chose for the XAFS studies were: 
as-deposited at
750~K, annealed at 1000~K and annealed at 1200~K. The annealed samples were
held at their respective temperatures for 10 hours in flowing oxygen, and
were heated and cooled at 2~K per minute.

The XAFS experiments were done on beamline 10-2 at SSRL using Si $<$220$>$
monochromator crystals and a 13-element Ge detector to collect
Mn $K_\alpha$ fluorescence data. The thin film samples were aligned at 
$\sim$55.0$^\circ$
with respect to the X-ray beam to make x, y and z axes equivalent, and thus
correspond to a powder. This angle comes from
the polarization dependence of the photoelectric effect\cite{Pettifer90}.
For each sample, we made
four sweeps at each temperature; for two of these sweeps, we rotated the
sample by 1.5$^\circ$ in order to determine the position of glitches, which
must be removed.

\section{Magnetization and Resistivity}

The magnetization $vs$ temperature data for these thin film samples are 
shown in Fig. \ref{magnetization}. All samples have broad 
ferromagnetic-to-paramagnetic phase 
transitions and T$_c$ increases with the annealing temperature. From these 
measurements, we extract T$_c$ for these samples (see Table I). 

The saturated magnetization of the samples increases with annealing temperature.
This indicates that at higher annealing temperature, a larger fraction of
the sample becomes ferromagnetic in the FM phase at low temperature. Our 
previous experiments show that the 30\% Ca doped LCMO powder sample has a Curie
temperature of $\sim$ 260~K\cite{Schiffer95}, which is very close 
to The T$_c$ of the fully annealed thin film sample.

Fig. \ref{rho} is a plot of log$_e$ of the resistivity $vs.$ temperature
for all three samples. We show both the data in zero field and
at 5.5 Tesla. 
The resistivity decreases with annealing temperature. At zero field,
a large peak is present for the 1000~K annealed sample; the resistivity
decreases and the peak moves to higher temperature for
the 1200~K annealed sample. There is no metal-to-insulator phase transition for
the as-deposited sample; this sample also shows a very large resistivity at low
temperature. When the external magnetic field is raised to 5.5 Tesla,
the resistivity drops
dramatically with magnetic field for the 1000~K and 1200~K annealed samples
at temperatures near the resistivity peak, and the
resistivity peaks move to higher temperature. For the as-deposited
sample, the biggest change in MR occurs at the lowest measuring temperature 
(70~K). It should also be noted that the peak in resistivity is very close to
T$_c$ for the 1200~K annealed sample, but well below T$_c$ (over 10~K
difference) for the 1000~K annealed sample.

Fig. \ref{mr} shows CMR\% (expressed as a percentage) $vs.$ temperature data 
for these samples on a log$_e$
scale. Here, we define CMR\% to be:

\vspace*{0.2in}
CMR\% = 100$\cdot$( R$_0$ - R$_H$ )/R$_H$
\vspace*{0.2in}

\noindent{where R$_0$ is the resistivity without magnetic field and R$_H$ is
the resistivity with a 5.5~T field.}

The magnetoresistance for these thin films is very large compared to that of
the corresponding powder samples\cite{Booth98b}. This is especially 
true for the as-deposited sample
at low temperature---the change of magnetoresistance at 70~K is about 3200\%.
The maximum CMR\% is about 1000\% for
the 1000~K sample and about 250\% for the 1200~K sample.

There is a double peak structure for the 1200~K annealed 
sample clearly visible in Fig.~\ref{mr}. As mentioned earlier,
our thin film samples are deposited on SrTiO$_3$ (STO) substrates. When the 
sample is deposited on STO, some Sr diffuses into the first few hundred \AA~ 
of the film and changes the stoichiometry
of that layer. This Sr diffusion could produce a resistance peak at a lower
temperature and may be responsible for this extra peak. We chose to use STO 
substrates instead of LaAlO$_3$ (LAO) 
substrates, which have no diffusion problem, because the large La 
$\it{L_I}$-edge XAFS
from the substrate would interfere with our Mn K edge XAFS data. Our XAFS 
data are sensitive to this Sr diffusion layer only if we focus on the 
further neighbors such as Mn-La, Mn-Ca and Mn-Sr. The Mn-O bond distance does
not change much between Ca and Sr substitution. Consequently for the Mn-O
pair distribution function which is the focus of this paper, we are not 
very sensitive to Sr in the lower 10\% of the film.

\section{XAFS Data Analysis and Discussion}

We collected all Mn-$K$ edge data in fluorescence mode. First the pre-edge
absorption (absorptions from other edges) was removed using the
Victoreen formula. Next we extract k$\chi$(k) where the photoelectron 
wave vector k is obtained from
k=$\protect\sqrt{2m_e(E-E_0)/h^2}$ and the XAFS function $\chi$(k)
is defined as $\chi$(k)=$\mu$(k)/$\mu_0$(k)-1. We fit a 7-knot spline to
$\mu$(k) (the $K$-edge absorption data above the edge) to obtain the 
background function $\mu_0$(k) (embedded atom function).
An example of such data
is shown in Fig. \ref{kspace}. Next we obtain the r-space data from the
Fourier Transform (FT) of k$\chi$(k); fits to the data were carried out 
in r-space\cite{Hayes82}. 
Some details of these fits are shown later in this paper.
See references \onlinecite{Booth98a,Booth98b,Rehr94}
for additional details.

In Fig. \ref{rspace}, we show the Mn $K$-edge Fourier transformed (FT) 
$r$-space data for
all three thin-film samples. In this figure, the position of each peak 
corresponds to an atom pair shifted by a well understood phase shift, 
$\Delta$r; for example, 
the first peak corresponds to the
Mn-O pairs and the second peak, which is near 3 \AA{}, 
corresponds to the Mn-La 
(Mn-Ca) pairs. The width, $\sigma$, of the pair distribution function is a 
measure of the local distortions in a shell of neighboring atoms. In XAFS,
a large $\sigma$ leads to a decrease in the amplitude of the $r$-space peak.
We obtain some results from these data without having to resort to curve
fitting. First, the amplitude of the Mn-O peak (the first peak) decreases
with increasing temperature. That means there are increasing distortions in 
each sample with increasing temperature.
Second, for the higher annealing temperature, the 
amplitude at low temperature increases and the
change of the amplitude with temperature for the Mn-O peak is larger than 
that for the other samples. The amount of distortion removed at low
temperature is smaller for the 1000~K annealed sample, 
while most of the distortion in the
as-deposited sample is still present at $T$ = 20~K. The amplitude of the r-space
data at high temperature (310~K) is almost the same for all three samples;
this suggests that the amount of distortion at high temperatures
doesn't change very much with the annealing temperature; however it is still
clearly less than the local distortion observed previously in LaMnO$_3$ at
300~K\cite{Booth98a,Booth98b}.

Our r-space data were fit using a Gaussian pair distribution function for the
Mn-O, Mn-Ca/La and Mn-Mn shells. The pair-distribution width,
$\sigma$(T), (for Mn-O) was determined from these detailed fits 
to the data, which were carried out using FEFF6 theoretical
functions\cite{Zabinsky95} (See Fig. \ref{s2}). In the fit, S$_0^2$N was 
fixed at 4.3, where N is the
number of nearest neighbors (6 O neighbors), and  S$_0^2$ 
is an amplitude reduction factor. k$\chi$(k) arrives from single-electron
process and is normalized to the step height; S$_0^2$ corrects for the fact
that the measured step height also includes multi-electron process. 
There is an absolute uncertainty in S$_0^2$ of roughly 10\%, however small 
changes in S$_0^2$ move all the curves up or down in Fig. \ref{s2} and do not 
change the shape or relative position.

The values of $\sigma^2$ obtained from the fits provide a measure of the
distortion of the Mn-O bond. Different contributions to the broadening add in
quadrature and hence $\sigma^2$ has the general form:

\[
\sigma^2 = \sigma^2_{static} + \sigma^2_{phonons} + \sigma^2_{other-mechanisms}
\]

At low temperature, $\sigma^2$ is dominated by zero-point motion and some 
static distortions. For all the manganites, the smallest value for $\sigma$ is 
about 0.04 \AA{}; consequently small variations in bond lengths, such as 
occurring for CaMnO$_3$ are not directly observed in $\sigma$. It is surprising
that the net distortion of the substituted CMR samples is about as small as
that observed for the more ordered CaMnO$_3$ structure at 20~K.

For each temperature 4 traces are analyzed
and averaged. The relative errors shown in Fig. \ref{s2} are the 
root mean square (rms) variation
of the fit result at each temperature. For the as-deposited sample,
$\sigma^2$ has a very small change with decreasing temperature; there is a 
larger change for the 1000~K annealed sample, and an even larger 
temperature dependence for the 1200~K annealed sample.
We also find in Fig. \ref{s2} that, at 320~K, $\sigma^2$
increases as the annealing temperature is lowered. This indicates that some of 
the static defects in the as-deposited sample can be removed during the 
annealing process.

The solid line in Fig. \ref{s2} corresponds to the data for the Mn-O bond 
in CaMnO$_3$, which has a high
Debye temperature (950~K); $\sigma^2$ for this sample will be denoted
$\sigma_T^2$ (The CaMnO$_3$ data we use in this 
figure are obtained from 
a powder sample\cite{Booth98b}. There might be a difference up to 
10\% between powder samples
and thin film samples since they are in different form and this difference 
may change the effective value of S$_0^2$ for the fit process.)

The difference between $\sigma^2_{data}$ and $\sigma^2_T$ at low temperature
is due to a static distortion $\sigma^2_{static}$, which is defined as:

\[
\sigma^2_{static} = \sigma^2_{data}(20K) - \sigma^2_T(20K).
\]

To estimate this quantity, we
shift the solid line vertically until it fits the low temperature data for
the 1200~K annealed sample This yields the dashed line in Fig. \ref{s2}, which
is defined to be $\sigma^2_T$ + $\sigma^2_{static}$.
Although $\sigma^2_{static}$ is almost zero for the 1200~K annealed sample,
it is large for the other samples. We include it here for the 1200~K annealed
sample to clarify its definition.
The contribution to $\sigma^2_{data}$ above the dashed line is attributed to a 
polaron distortion, where the full (maximum) polaron distortion, 
$\sigma^2_{FP}$, is defined
by:

\[
\sigma^2_{FP} = \sigma^2_{data}(300K) - \sigma^2_T(300K) - \sigma^2_{static}
\]

We have found in previous work\cite{Booth98a,Booth98b} that a useful 
parameter is the distortion removed as T drops below T$_c$, $\Delta\sigma^2$,
which we define below. First,
the $\sigma^2_T$ curve is shifted vertically (by an amount
$\sigma^2_{FP}$ + $\sigma^2_{static}$) such that it
fits the high temperature data. This yields the dotted line shown in 
Fig. \ref{s2} which is $\sigma^2_T$ + $\sigma^2_{static}$ + $\sigma^2_{FP}$.
This dotted line represents the expected Debye behavior plus static 
distortion if no polaron distortion were removed upon cooling.
We define $\Delta\sigma^2$ as the difference 
between the dotted line and the data:

\[
\Delta\sigma^2 = \sigma^2_T + \sigma^2_{FP} + \sigma^2_{static} - \sigma^2_{data}
\]

A similar analysis is carried out for the as deposited
sample and the 1000~K annealed sample (corresponding curves for $\sigma_T^2$
+ $\sigma_{static}^2$ and
$\sigma_T^2$ + $\sigma_{static}^2$ + $\sigma_{FP}^2$ are not shown in 
Fig. \ref{s2}). It is also important to point out that the
difference between $\sigma^2$ for the 1200~K annealed sample and that of
CaMnO$_3$ at 20~K is very small, which suggests that the Mn-O local structure 
of the fully annealed sample can be as ordered as that of CaMnO$_3$ even though
it is a doped sample. The same
result was obtained for La$_{1-x}$Ca$_x$MnO$_3$ CMR powder samples 
(x = 0.2$\sim$0.5) from our previous experiments\cite{Booth98a,Booth98b}.
The reason for this phenomena for the thin films can be explained as follows:
first, the high temperature annealing process appears to remove most of the
static defects such as dislocations and vacancies; second, at very low 
temperatures, there is almost no difference
between the two type of Mn sites in the DE model, the electron moves 
rapidly from one site to another on a time scale fast compared to the 
appropriate phonon
frequency. Consequently, Jahn-Teller distortions don't have time to form.

For the as-deposited sample, the large CMR\% occurs when 
there is a large value for $\sigma^2$ at low temperature; this suggests that
the distortion in the sample may, in part, be the origin of the unusually large 
magnetoresistance in thin-film samples. We have recently observed similar
results in our study of Ti and Ga doped LCMO powder samples\cite{Dcao99}.

In Fig. \ref{lns2}, we plot $\ln \Delta\sigma^2$ {\it vs} $M/M_0$.
Our previous studies
of La$_{1-x}$Ca$_x$MnO$_3$ powder samples showed that there is a linear
relationship between $\ln \Delta\sigma^2$ and the
magnetization\cite{Booth98a,Booth98b}, which
provides evidence that there is a strong connection between the local distortion
and magnetism in these materials. The solid squares in Fig. \ref{lns2} show
the linear
relationship for a La$_{0.70}$Ca$_{0.30}$MnO$_3$ powder sample from our previous
work\cite{Booth98a,Booth98b}. For the thin film samples, there is a similar 
connection
between local distortion and magnetization.
For the 1000~K and 1200~K annealed samples, we find a small deviation from
a straight line below M/M$_0$ $\sim$0.3. Since the error of the data in this
range (M/M$_0$ $<$ 0.3) is large, it is not clear
if this is a real effect or not. The data for the as-deposited sample
appear to follow a straight line, but the errors in the difference are too
large to draw conclusions. Also, we find that the data for the 1200~K annealed
thin-film and the powder sample (both La$_{0.70}$Ca$_{0.30}$MnO$_3$) almost 
overlap each other. This suggests that the local structure of the fully annealed
thin-film behaves similarly to that of the corresponding bulk sample.

In order to see the effect of annealing temperature on the local distortion more
clearly, we plot the distortion contributions, $\sigma_{static}^2$ and 
$\sigma_{FP}^2$, as a function of annealing temperature in
Fig. \ref{disorder}. This figure shows that the static
distortion decreases with annealing temperature, while the polaron contribution 
increases with annealing temperature. This suggests that part of the static 
distortions observed in the as-deposited sample become dynamic, polaron-like,
distortions after annealing, for T $>$ T$_c$. 

It should be noted that although the magnetization
only drops by roughly 50\% for the as-deposited sample, the $\sigma_{FP}^2$ 
contribution becomes much smaller ( of order 5\%). Consequently, there must be 
statically
distorted regions in the as-deposited sample that are also ferromagnetic.
The reduction of the saturated magnetization can arise in several ways.
First, because of the inhomogeneous material, small regions may be 
antiferromagnetic
(AF). Second, it has been suggested that the crystal field, particularly 
in regimes where
the inhomogeneity causes a local reduction of the tolerance factor, can result
in a low spin Mn ($^4$t$_{2g}$) configuration with a local 50\% decrease in 
Mn moment\cite{Geballe98}. Third, the spins in a small domain may not be 
exactly parallel.
However, to explain the entire decrease in saturated magnetization would require
very large canting angles. Fourth, the magnetization vectors of each domain
may not be aligned. The number of domain-walls may be important for calculating
the resistivity. However, for domains large compared to a unit cell, slightly
canted spins within a domain or a lack of alignment of the magnetization of 
various domains would not lead to a significant decrease in the polaron 
contribution to the broadening in XAFS. Consequently, the presence of static 
distortions but essentially no polaron-like contributions in the as-deposited
material indicate both a significant fraction of AF material, and the 
presence of some low spin Mn sites, possibly induced by the disorder 
(interstitials, vacancies, inhomogeneous Ca concentrations etc.).
This disorder must pin the local distortions which would also suppress the 
electron
hopping frequency, thereby reducing the effectiveness of the DE interaction.

All three thin film samples were prepared in the same way, except for the
annealing temperature. During the annealing process, part of the static
defects in the sample such as vacancies and interstitials can be removed.
In addition, the annealing process can also change the amount of oxygen in
the sample\cite{Worledge96,Worledge98,Ju95,Malde98}. The sample is slightly
oxygen deficient before the annealing process, and oxygen is
incorporated during annealing. The fully annealed sample
(1200~K) is expected to be fully
stoichiometric ( similar to the corresponding powder sample).
It is well documented
that samples without sufficient oxygen can have higher resistivities, a lower
resistance peak temperature, a lower T$_c$ and a lower saturated
magnetization\cite{Worledge96,Worledge98,Ju95,Malde98}.
We have the same trends in our experiments.

Compared to the 1200~K annealed sample,
the samples annealed at lower temperatures have a large decrease in 
$\sigma^2_{FP}$
and a large increase in resistivity, while the saturated magnetization changes
by less than a factor of two. This suggests that much of the distorted 
magnetic regions 
probably do not contribute to the conductivity and that the fraction of 
conducting material is very low for the as-deposited sample. However, the six
order of magnitude increase in resistivity is much larger than the volume 
reduction of the regions that still have a polaron-like distortion. 
Consequently, it is likely that percolation also plays a role for transport in 
the as-deposited  sample. Then the magnetic field may play two roles for this 
sample; it will decrease the resistivity for the conducting pathways that 
exist at B=0 and may also make some "marginal" pathways become conducting.

\section{Conclusion}

From our analysis, we find that the annealing temperature of the thin films
affects the local distortion of the materials appreciably.
The large change in resistivity and the small change in local structure with
temperature for the
as-deposited sample suggest that only small regions are contributing to the
resistivity and percolation may play a role. We also find that
there is still a strong connection between local distortions, resistivity and
magnetism in the fully annealed thin-film materials, which behaves much 
like a powder sample. For the 1000~K annealed sample, the resistivity and MR
peaks are well below T$_c$. In this case (including the as-deposited sample),
the local distortions correlate well with the magnetization, but there is no 
feature in $\sigma$ that occurs at the temperature at which the resistivity 
has a peak.

\acknowledgements{
This work was supported in part by NSF grant DMR9705117.
The experiments were performed at SSRL, which is operated by the DOE,
Division of Chemical Sciences, and by the NIH, Biomedical Resource Technology
Program, Division of Research Resources.}

\begin{table}
\label{Tc}
\caption{The Curie temperature for each thin
film sample.}
\vspace*{0.1in}
\begin{tabular}{|l|c|c|c|}
 sample   & As-deposited     & Annealed 1000K     & Annealed 1200K \\
\hline
 T$_c$ (K)& 164(5)           & 202(5)             & 257(5)         \\
\end{tabular}
\end{table}

\begin{figure}
\vspace*{0.15in}
\caption{A plot of magnetization $vs$ temperature for
La$_{0.67}$Ca$_{0.33}$MnO$_3$ thin film samples, with an applied field of
 H=5000~Oe.}
\label{magnetization}
\end{figure}

\begin{figure}
\vspace*{0.15in}
\caption{This plot shows the resistivity data (ln scale) for three thin film
samples, with and without magnetic field. The open symbol corresponds
to data without field and the solid symbol corresponds to the data with a
field of 5.5~T.
There is a double peak structure
for the 1200~K annealed sample around 150~K, which is just visible in this
figure.
}
\label{rho}
\end{figure}

\begin{figure}
\vspace*{0.15in}
\caption{A plot of CMR\% (ln scale) $vs$ temperature for
La$_{0.67}$Ca$_{0.33}$MnO$_3$
thin film samples in a field of 5.5~T. The double peak structure
for the 1200~K annealed sample is obvious in this plot.
}
\label{mr}
\end{figure}

\begin{figure}
\vspace*{0.15in}
\caption{A plot of the $k$-space data for the 1200~K annealed sample at 20~K
to show the quality of the data. Although there is some noise in the data, the
quality is good up to 11 \AA$^{-1}$
}
\label{kspace}
\end{figure}

\begin{figure}
\vspace*{0.15in}
\caption{A comparison of the change in the FT, $r$-space, data with temperature
for the La$_{0.67}$Ca$_{0.33}$MnO$_3$ thin-film samples with different annealing
temperatures. Top: 750~K anneal (as deposited), middle: 1000~K anneal, bottom:
1200~K anneal. FT range is 3.3--10.5~\AA$^{-1}$, with 0.3~\AA$^{-1}$ Gaussian
broadening. The curve inside the envelope, with a higher frequency, is the
real part of the FT (FT$_R$).
The envelope is defined as:
$\pm \protect\sqrt{FT_R^2 + FT_I^2}$
where FT$_I$
is the imaginary part of the FT.
}
\label{rspace}
\end{figure}

\begin{figure}
\vspace*{0.15in}
\caption{A plot of $\sigma^2$ $vs$ temperature for the as-deposited,
1000~K and 1200~K annealed La$_{0.67}$Ca$_{0.33}$MnO$_3$ thin-film samples.
(Here $\sigma$ is the pair-distribution width of the Mn-O peak)
The solid line is the thermal contribution $\sigma_T^2$ from 
CaMnO$_3$ \protect\cite{Booth98a,Booth98b}
the dashed line corresponds to $\sigma_T^2$ + $\sigma_{static}^2$ and the
dotted line is $\sigma_T^2$ + $\sigma_{static}^2$ + $\sigma_{FP}^2$ for the
annealed 1200~K sample (see text).}
\label{s2}
\end{figure}

\begin{figure}
\vspace*{0.15in}
\caption{A plot of ln$\Delta\sigma^2$ $vs$ $M/M_0$ is presented
for the as-deposited, 1000~K,
and 1200~K annealed La$_{0.67}$Ca$_{0.33}$MnO$_3$ thin-film samples, as well as
a La$_{0.70}$Ca$_{0.30}$MnO$_3$ powder sample. M is the
measured magnetization and M$_0$ is the saturated magnetization;
M/M$_0$ is the relative magnetization.}
\label{lns2}
\end{figure}

\begin{figure}
\vspace*{0.15in}
\caption{Static and polaron distortion $vs$ annealing temperature for all
three thin film samples. The solid triangle symbol represents the static
distortion, $\sigma^2_{static}$; the open square symbol represents the
full-polaron distortion, $\sigma^2_{FP}$.}
\label{disorder}
\end{figure}


\begin{thebibliography}{10}

\bibitem{Zener51}
C. Zener, Phys. Rev. {\bf 82}, 403 (1951).

\bibitem{Anderson55}
P.~W. Anderson, H. Hasegawa, Phys. Rev. {\bf 100}, 675 (1955).

\bibitem{deGennes60}
P.~G. de~Gennes, Phys. Rev. {\bf 118}, 141 (1960).

\bibitem{Millis95}
A.~J. Millis, P.~B. Littlewood, and B.~I. Shraiman, Phys. Rev. Lett. {\bf 74},
  5144  (1995).

\bibitem{Booth98a}
C.~H. Booth, F. Bridges, G.~H. Kwei, J.~M. Lawrence, A.~L. Cornelius, and J.~J.
  Neumeier, Phys. Rev. Lett. {\bf 80},  853  (1998).

\bibitem{Booth98b}
C.~H. Booth, F. Bridges, G.~H. Kwei, J.~M. Lawrence, A.~L. Cornelius, and J.~J.
  Neumeier, Phys. Rev. B {\bf 57}, 10440 (1998).

\bibitem{Billinge96}
S.~J.~L. Billinge, R~.G. DiFrancesco, G.~H. Kwei, J.~J. Neumeier, and J.~D. 
Thompson, Phys. Rev. Lett. {\bf 77}, 715 (1996).

\bibitem{Louca97}
D. Louca, T. Egami, E.~L. Brosha, H. R{\"{o}}der, and A.~R. Bishop,
Phys. Rev. B {\bf 56}, R8475 (1997).

\bibitem{Worledge96}
D.~C. Worledge, G.~Jeffrey Snyder, M.~R. Beasley, and T.~H. Geballe,
J. Appl. Phys. {\bf 80}, 5158 (1996). 

\bibitem{Worledge98}
D.~C. Worledge, L. Mieville, T.~H. Geballe,
J. Appl. Phys. {\bf 83}, 5913 (1998).

\bibitem{Ju95}
H.~L. Ju, J.~Gopalakrishnan, J.~L. Peng, Qi~Li, G.~C. Xiong, T.~Venkatesan 
and R.~L. Greene, Phys. Rev. B {\bf 51}, 6143 (1995). 

\bibitem{Malde98}
N. Malde, P. S. I. P. N. De Silva, A. K. M. Akther Hossain, L. F. Cohen,
K. A. Thomas, J. L. MacManus-Driscoll, N. D. Mathur and M. G. Blamire,
Solid State Comm. {\bf 105}, 643 (1998).

\bibitem{Norby95}
P. Norby, I.~G. {Krogh Andersen}, E. {Krogh Andersen}, and N.~H. Andersen, J.
  Solid State Chem. {\bf 119},  191  (1995).

\bibitem{Mitchell96}
J.~F. Mitchell, D.~N. Argyriou, C.~D. Potter, D.~G. Hinks, J.~D. Jorgensen, and
  S.~D. Bader, Phys. Rev. B {\bf 54},  6172  (1996).

\bibitem{Poeppelmeier82}
K.~R. Poeppelmeier, M.~E. Leonowicz, J.~C. Scanlon, and J.~M. Longo, J. Solid
  State Chem. {\bf 45},  71  (1982).

\bibitem{Pettifer90}
R.F. Pettifer, C. Brouder, M. Benfatto, C.~R. Natoli, C. Hermes,
M.~F. Ruiz Lopez, Phys. Rev. B {\bf 42}, 37 (1990).

\bibitem{Schiffer95}
P. Schiffer, A. Ramirez, W. Bao, and S-W. Choeng, Phys. Rev. lett. {\bf 75},
3336 (1995).

\bibitem{Hayes82}
T. M. Hayes and J. B. Boyce, Solid State Physics {\bf 37}, 173 (1982).

\bibitem{Rehr94}
J.~J. Rehr, C.~H. Booth, F. Bridges, S.~I. Zabinsky,
Phys. Rev. B {\bf 49}, 12347 (1994).

\bibitem{Zabinsky95}
S.~I. Zabinsky, A. Ankudinov, J.~J. Rehr, R.~C. Albers,
Phys. Rev. B {\bf 52}, 2995 (1995).

\bibitem{Dcao99}
D. Cao, F. Bridges, A.~P. Ramirez, M. Olapinski, M.~A. Subramanian,
C.~H. Booth, G. Kwei, Unpublished.

\bibitem{Geballe98}
T.~H. Geballe, B.~Y. Moyzches, 5th International Workshop in Oxide
Electronics, Dec. 7 1998, U. of Maryland, Unpublished.

\end{thebibliography}
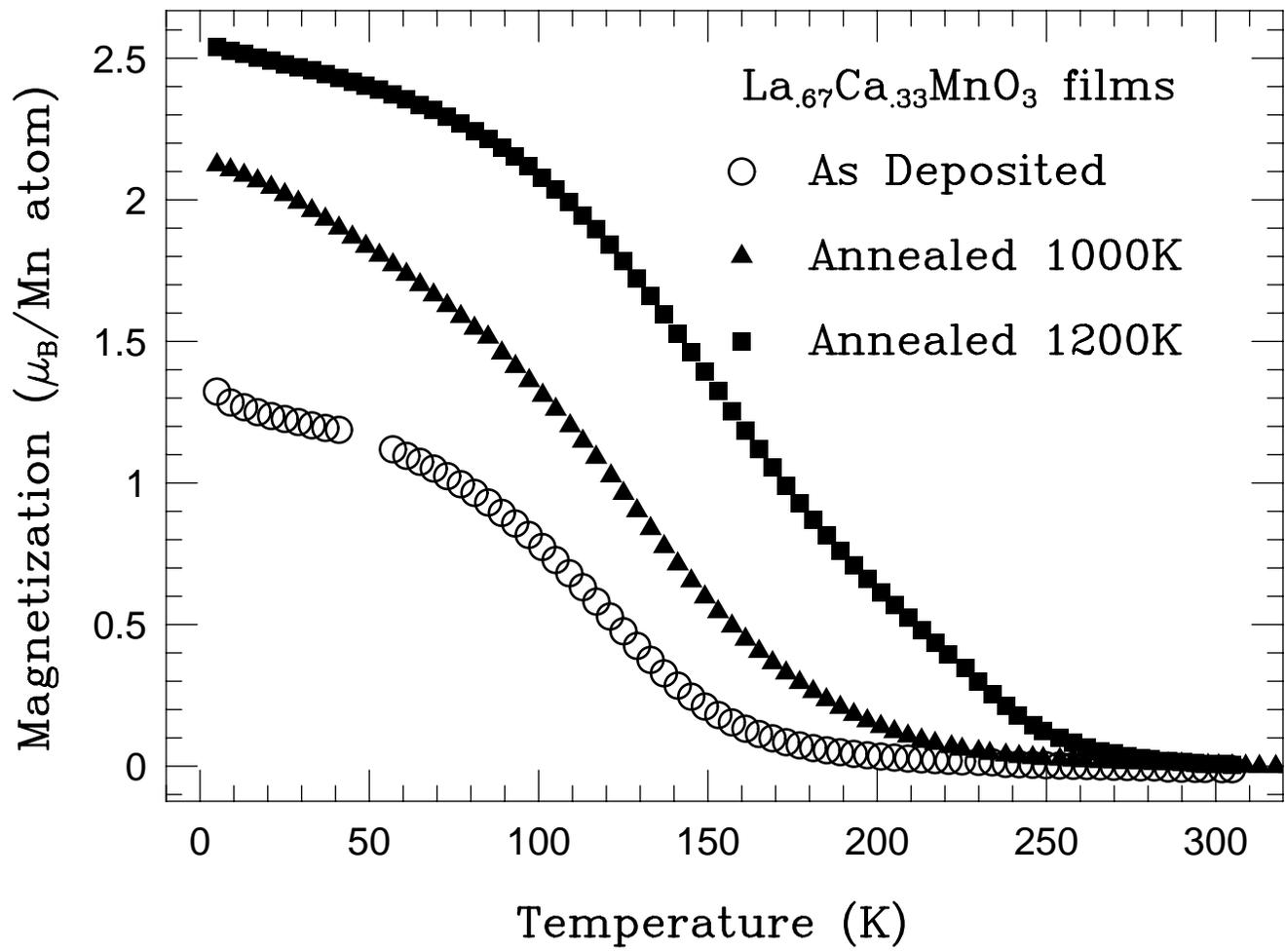
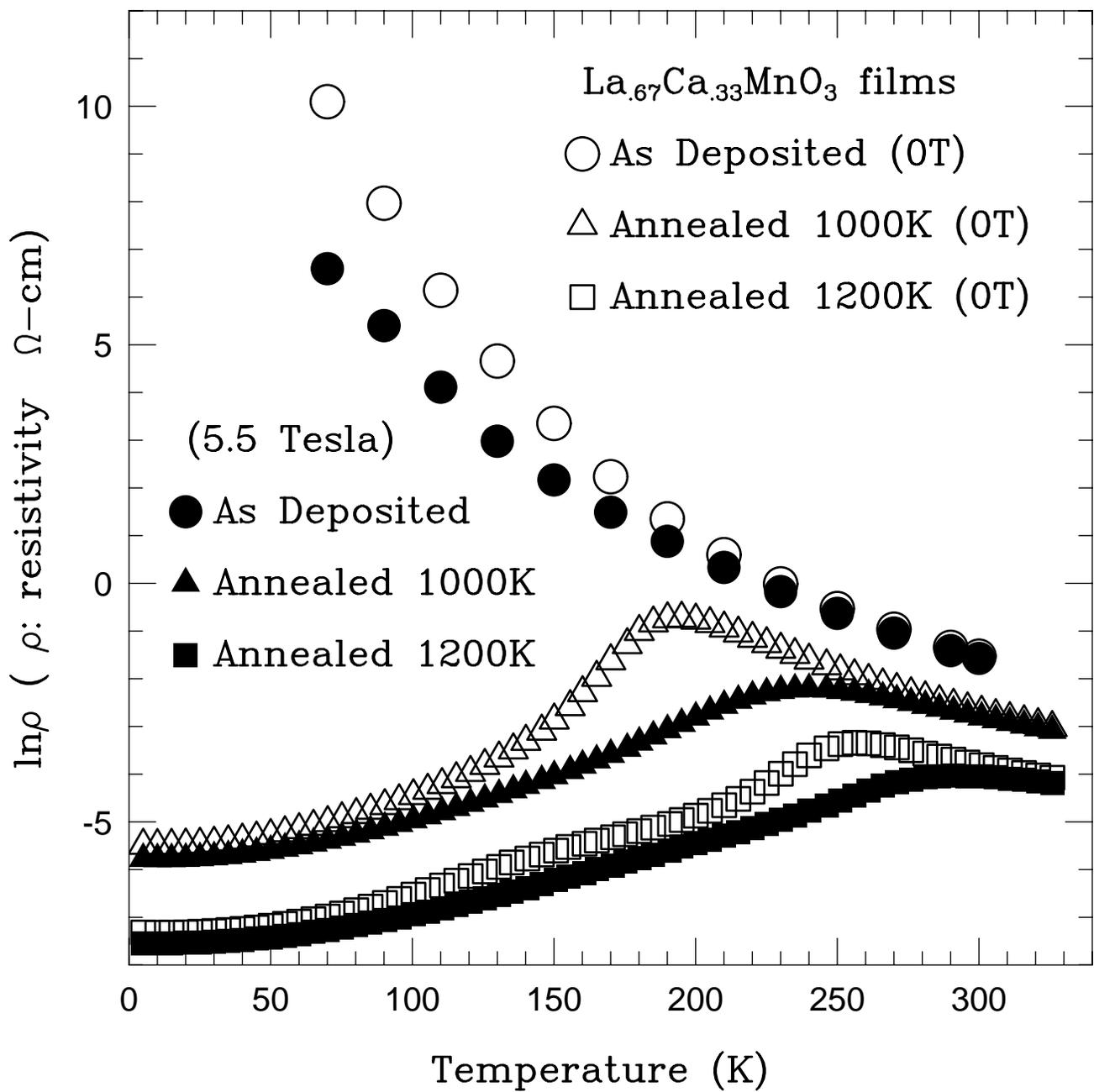
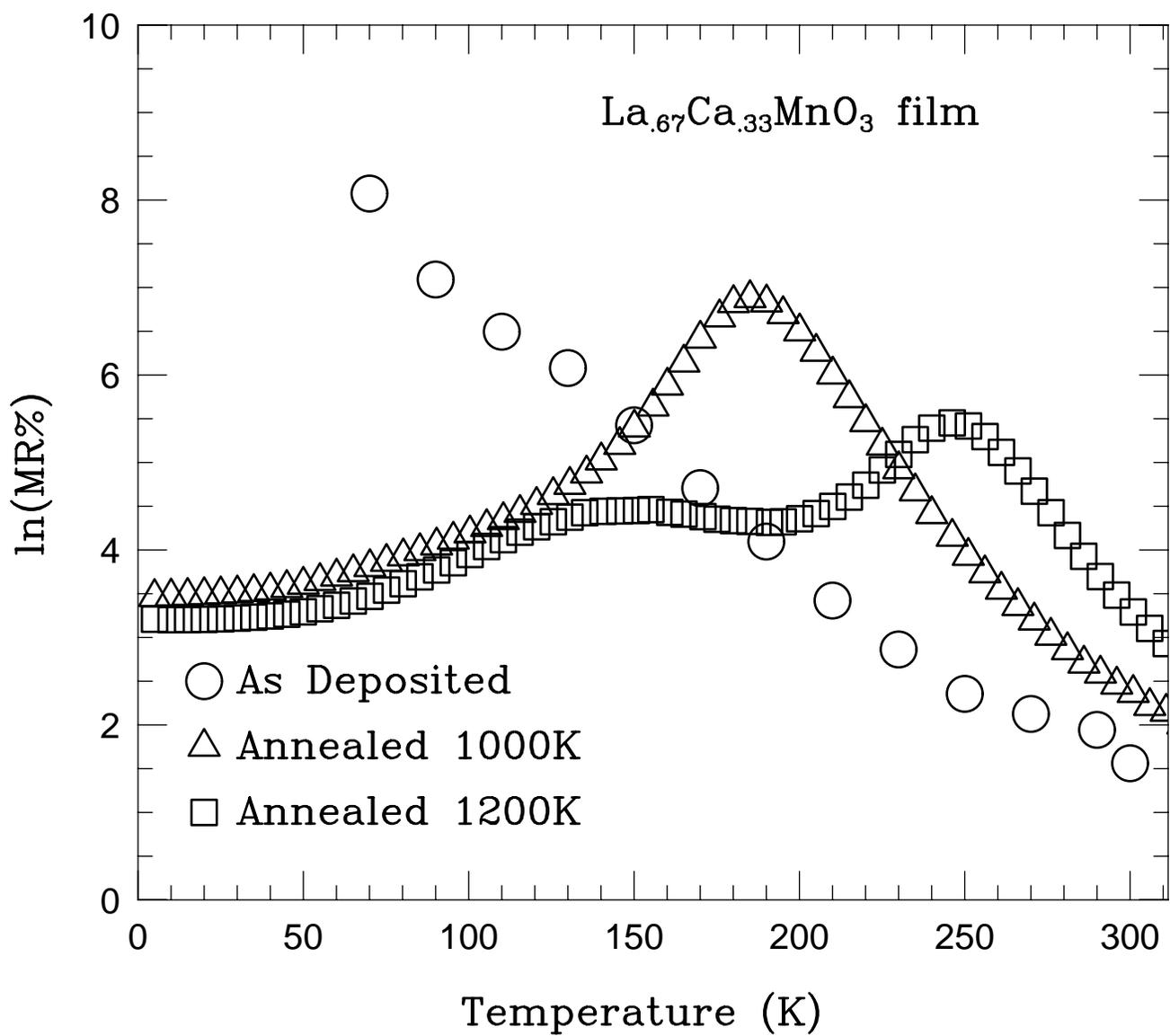
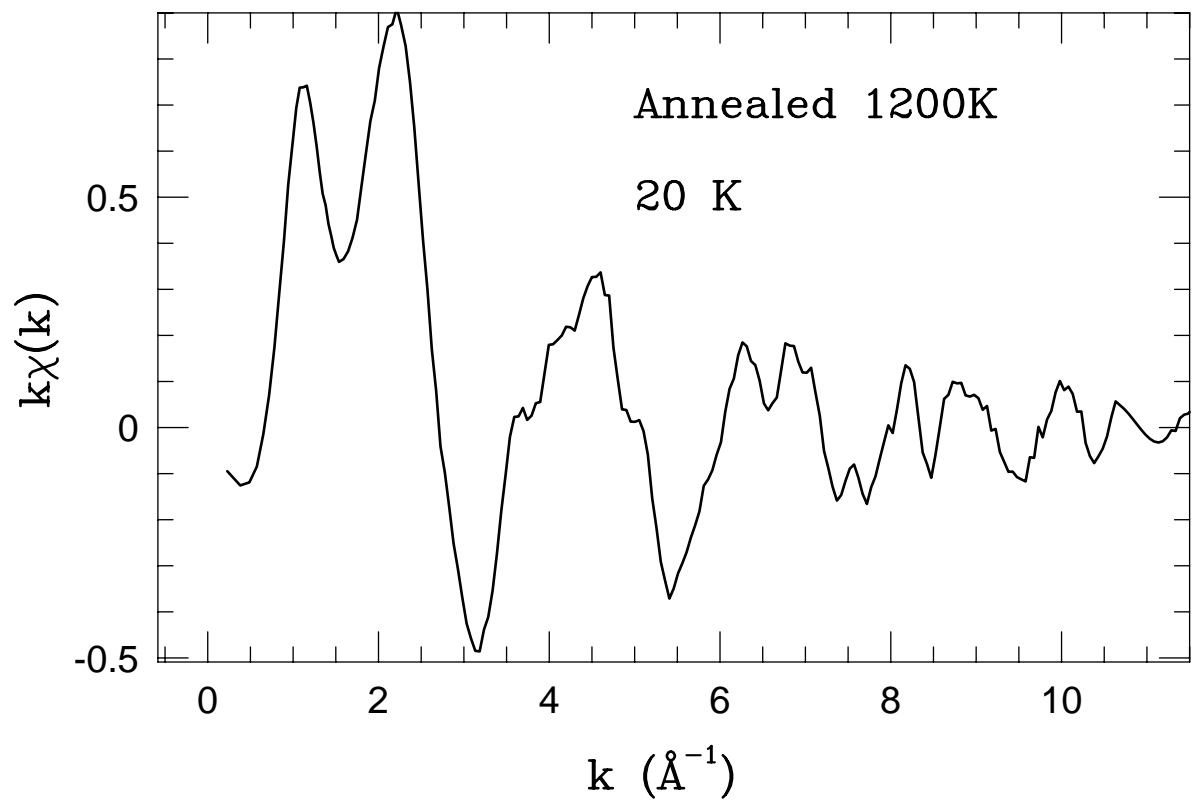
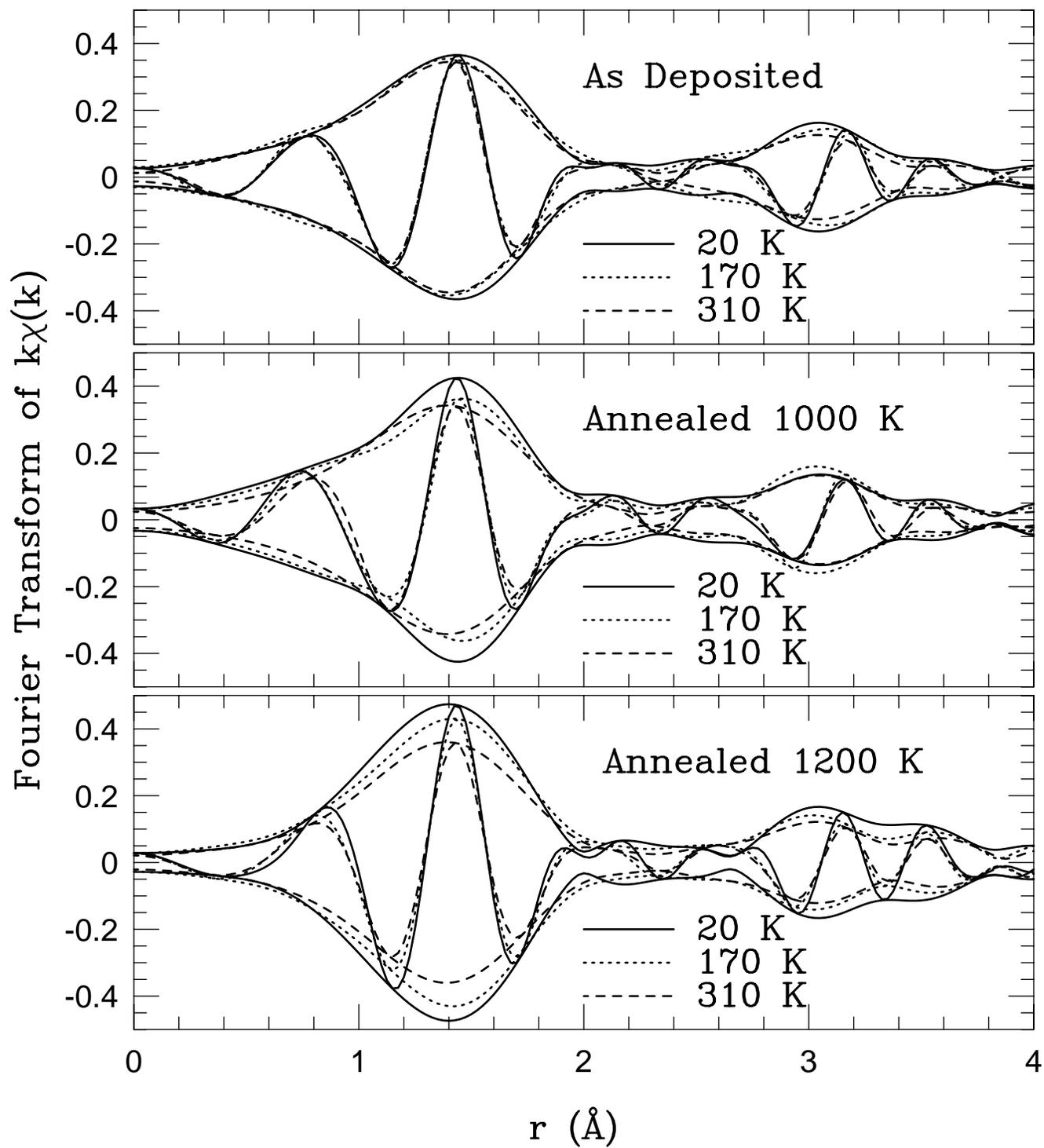
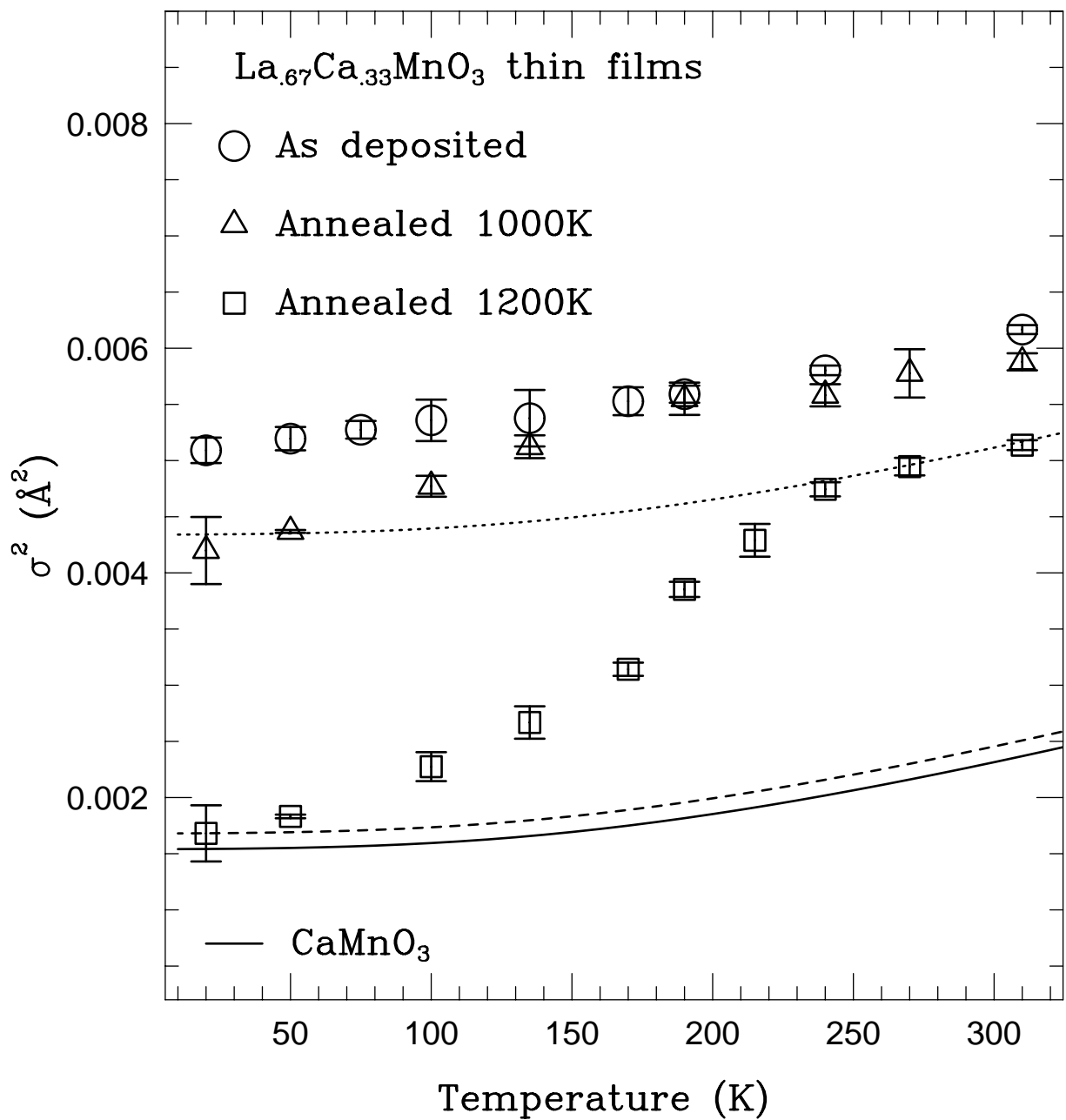
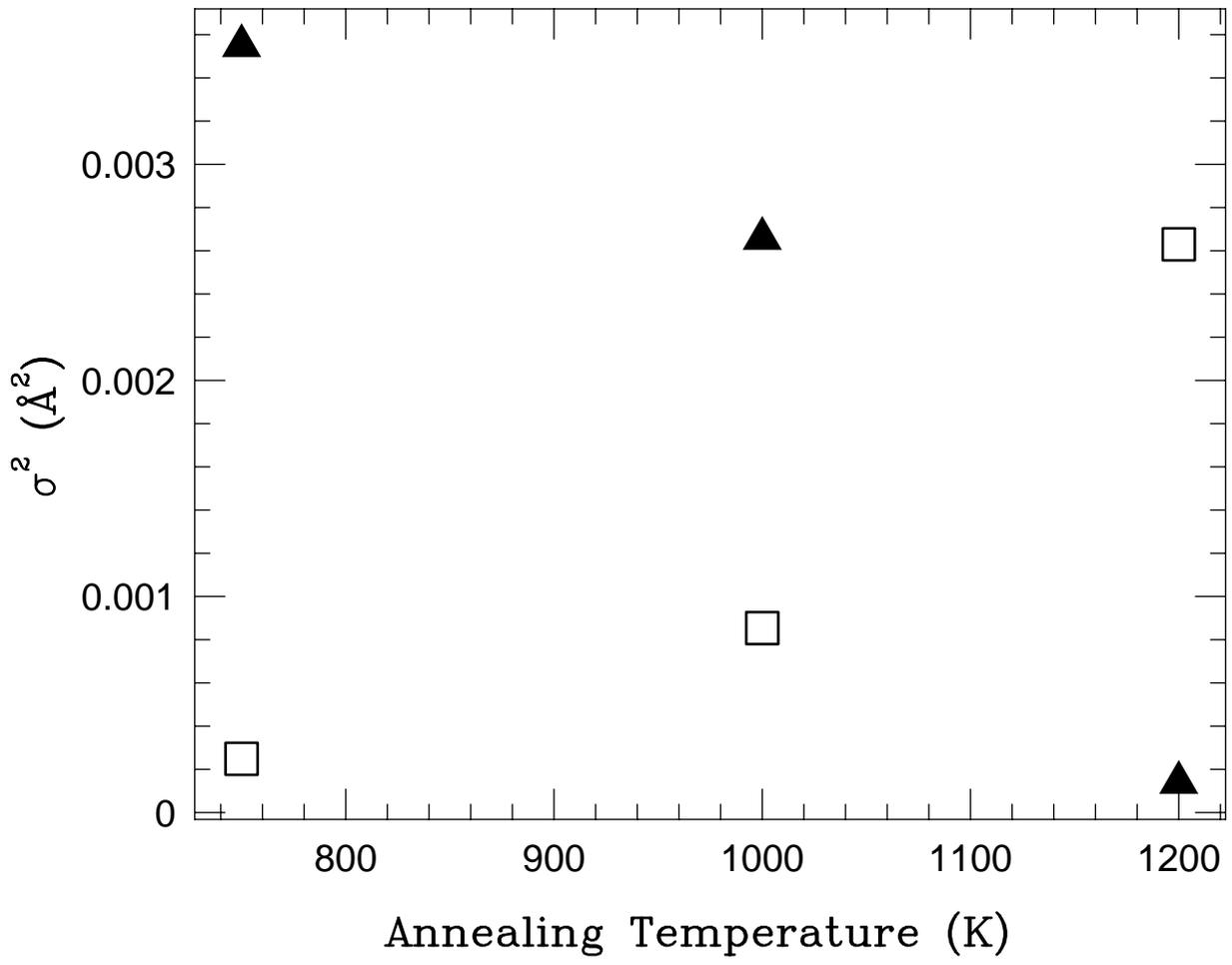
\end{document}